%% file: main.tex
\documentclass[conference, 10pt, a4paper]{IEEEtran}

\input{settings.tex}

\begin{document}

\sloppy

\title{An Energy-Efficient Event-Based MIMO Communication Scheme for UAV Formation Control}
\author{\IEEEauthorblockN{Yasemin Karacora, Aydin Sezgin}
 \IEEEauthorblockA{Institute of Digital Communication Systems, Ruhr University Bochum\\ Email: \{yasemin.karacora, aydin.sezgin\}@rub.de}}

\maketitle

\begin{abstract}
    We consider a leader-follower formation control setup as an example for a multi-agent networked control system (NCS). This paper proposes an event-based wireless communication scheme with a MIMO precoder, that dynamically adapts to both the channel and the current control state. We apply a Lyapunov drift approach for optimizing the precoder and for defining an event-triggering policy. To evaluate the performance of the proposed scheme, we simulate a formation control of two unmanned aerial vehicles (UAVs) connected via a point-to-point MIMO fading channel. Compared to a benchmark scheme with periodic transmissions and a basic water-filling algorithm, we demonstrate that our proposed event-based scheme is more efficient by consuming less energy and requiring less frequent transmissions on average. 
\end{abstract}

\section{Introduction}
\input{content/intro.tex}

\section{System Model}\label{sec:systemmodel}
\input{content/system_model.tex}

\section{Lyapunov drift optimization}\label{sec:optimization}
\input{content/optimization.tex}

\section{Simulation Results}\label{sec:simulation}

\input{content/simulation.tex}
\vspace{0.5em}
\section{Conclusion}\label{sec:conclusion}
\input{content/conlusion.tex}

\bibliographystyle{IEEEtran}
\bibliography{references}

\end{document}

%% file: settings.tex
\usepackage{standalone}
\usepackage[reqno]{amsmath}
\usepackage{amssymb}

\usepackage{algorithm}
\usepackage{algorithmicx}
\usepackage{algpseudocode}
\usepackage{setspace}
\usepackage{multicol}
\usepackage{enumerate}
\usepackage[font=small, belowskip=-15pt,aboveskip=0pt]{caption}
\usepackage{cite}
\usepackage{amsthm}
\usepackage[]{graphicx}
\usepackage{tikz}
\usepackage{tkz-tab}
\usepackage{pgfplots}
\usepackage{booktabs}
\usepackage{floatflt}

\usepackage{psfrag}	
\usepackage{array}
\usepackage{multirow,hhline}
\usepackage{exscale}
\usepackage{color}
\usepackage{colortbl}
\usepackage{epsfig}

\usepackage{array}
\usepackage[tight, nooneline]{subfigure}
\subfiguretopcaptrue

\usepackage[latin1]{inputenc} 
\usepackage[T1]{fontenc} 
\usepackage{rotating}
\usepackage{pbox}

\usepackage{pifont}

\usepackage{comment}

\newcommand{\mat}[1]{\ensuremath{\mathbf{#1}}}

\DeclareMathOperator{\tr}{Tr}

\DeclareMathOperator{\diag}{diag}

\newcommand{\ev}{{\bf e}}

\newcommand{\sv}{{\bf s}}

\newcommand{\uv}{{\bf u}}
\newcommand{\wv}{{\bf w}}
\newcommand{\vv}{{\bf v}}
\newcommand{\xv}{{\bf x}}
\newcommand{\yv}{{\bf y}}
\newcommand{\zv}{{\bf z}}

\newcommand{\Amat}{{\mat{A}}}
\newcommand{\Bmat}{{\mat{B}}}
\newcommand{\Cmat}{{\mat{C}}}

\newcommand{\Fmat}{{\mat{F}}}
\newcommand{\Gmat}{{\mat{G}}}
\newcommand{\Hmat}{{\mat{H}}}
\newcommand{\Imat}{{\mat{I}}}

\newcommand{\Kmat}{{\mat{K}}}

\newcommand{\Mmat}{{\mat{M}}}

\newcommand{\Pmat}{{\mat{P}}}
\newcommand{\Qmat}{{\mat{Q}}}
\newcommand{\Rmat}{{\mat{R}}}
\newcommand{\Smat}{{\mat{S}}}
\newcommand{\Tmat}{{\mat{T}}}
\newcommand{\Umat}{{\mat{U}}}
\newcommand{\Vmat}{{\mat{V}}}
\newcommand{\Wmat}{{\mat{W}}}
\newcommand{\Xmat}{{\mat{X}}}
\newcommand{\Ymat}{{\mat{Y}}}

\newcommand{\Zeromat}{{\mat{0}}}

\usetikzlibrary{shapes,snakes}
\usetikzlibrary{calc}
\usetikzlibrary{patterns}
\usetikzlibrary{decorations.pathmorphing} 
\usetikzlibrary{spy}
\usetikzlibrary{positioning}
\usetikzlibrary{arrows, decorations.markings}
\usetikzlibrary{shapes.multipart, fit, backgrounds}

%% file: content/intro.tex
The 5G mobile communication standard is the first that will be applicable to mission-critical control systems. 
One of three categories of 5G services is the ultra-reliable low-latency communications (uRLLC), which is designed to meet the stringent requirements of industrial automation applications as well as automated driving. In particular, using wireless networks is essential in multi-agent networked control systems (NCS) that comprise mobile objects, such as moving robots in industrial manufacturing or self-driving cars. In this type of systems, synchronization tasks arise in order to accomplish a common control goal or to avoid collisions between the agents. This leads to high demands on the Quality of Service (QoS) of the wireless network that have to be guaranteed despite limited energy resources at the mobile agents.
Hence, an energy-efficient yet stability-maintaining scheme is of great interest. An event-based approach, that initiates communication only when necessary, can generally reduce the amount of data to be exchanged among agents. In contrast to clock-based communication, where transmission takes place in periodic time intervals, the event-based approach enhances efficiency as it allows for a dynamic adjustment of the communication to the current state of the controlled system. Hence, resources can be saved while the system is in a non-critical state to be used whenever it is necessary to maintain stability or satisfy performance requirements.

While the majority of existing research considers the design of the communication and control schemes separately, communication and control co-design in NCSs has recently gained more interest, e.g. in \cite{cai2016mimo, eisen2019control, zeng2018integrated}. 
In particular, the authors in \cite{cai2016mimo} present an event-based MIMO precoding scheme for an NCS with an energy harvesting sensor. They propose a waterfilling solution that in addition to the channel state also takes account of the available energy and the plant state estimation error. However, the communication scheme is independent of the plant state itself.
Event-triggered communication has also been studied by the authors in \cite{schwung2019networked}, where a collision avoidance problem of two unmanned aerial vehicles (UAV) is considered, but their focus is on the control and event generator design.
The authors in \cite{santos2019self} present a self-triggered formation control scheme, in which a remote controller requests sensor measurements in an event-based fashion using Lyapunov control theory. 
A leader-follower formation control problem is considered in \cite{gao2018fixed} for autonomous underwater vehicles. The authors propose a scheme that switches between continuous and periodic communication on an event-triggered basis.

We aim at designing a communication scheme for a formation control application that ensures both energy efficiency and formation accuracy by dynamically adjusting to the current control state. For this purpose, we propose a control-aware MIMO precoder combined with an event-triggering strategy.
In this paper, we consider a leader-follower UAV formation control problem with one leading and one following agent. Although our control setup is different from \cite{cai2016mimo}, the Lyapunov optimization method proposed there proves useful for our precoder design as well. More specifically, a closed-form solution for the precoder is derived by minimizing a Lyapunov drift function subject to a transmit power constraint. 
However, while the focus in \cite{cai2016mimo} is on the control state estimation error only, we combine this with an event-triggering policy that additionally considers the current estimated control deviation. Intuitively, this is important since the same estimation error can be much more critical if the system already deviates from its desired state.
We evaluate the performance of our proposed scheme through simulations, getting insights into the benefits of such a control-aware communication design in terms of energy efficiency and control accuracy compared to a periodic transmission scheme with conventional water-filling that is designed independently of the control system. In particular, we demonstrate an improved energy efficiency both in terms of transmit power and control input power.

%% file: content/system_model.tex
\begin{figure*}
\centering
\setlength{\abovecaptionskip}{6pt}
\includegraphics[]{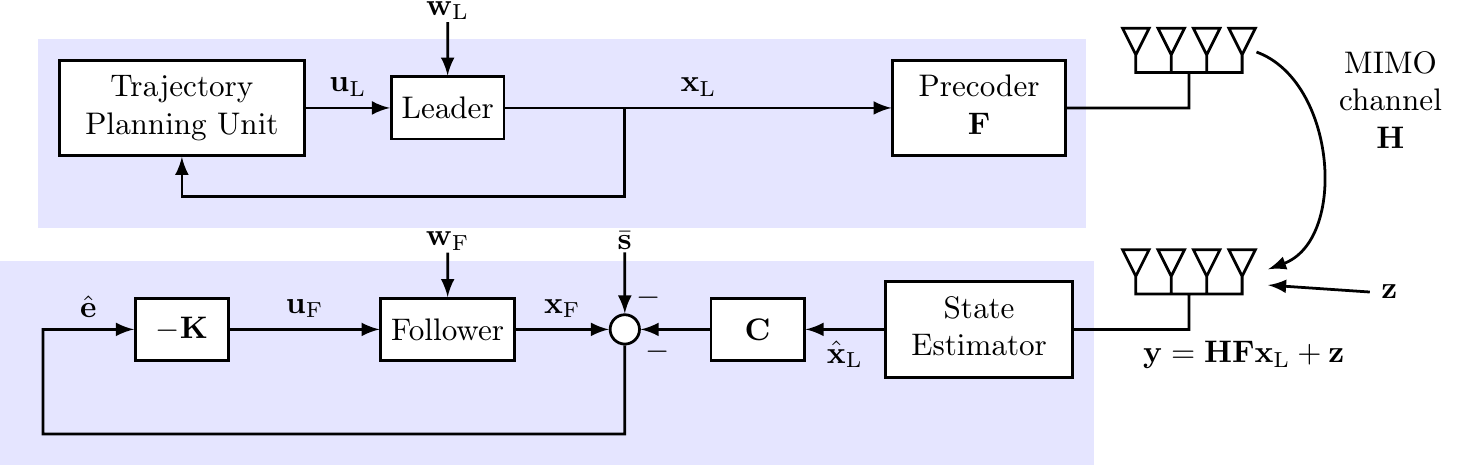}
    \caption{System model of the Leader-Follower formation control setup.}
    \label{system_model}
\end{figure*}

We consider a leader-follower formation control setup with one leader and one follower, that are communicating over a wireless network, as depicted in Figure \ref{system_model}. 
More precisely, while the leader follows a given path that is unknown to the follower, the follower is supposed to move alongside with a constant relative position vector denoted by $\Bar{\sv}$ between the two agents. The leader's current position is communicated to the follower wirelessly upon request. The communication takes place in an event-based fashion, i.e., communication is invoked on an event occurence, such as the control deviation exceeding a predefined threshold. 

The follower tracks the leader's state using a Kalman filter with intermittent observations as presented in \cite{sinopoli2004kalman}. Based on the estimated state of the leader, the follower's position is controlled. 

\subsection{Leader-Follower Formation Control Architecture}

The system dynamics of the leader and follower are modeled using a discrete-time linear state space representation as follows. The leader's dynamics are given by
\begin{equation} \label{x_leader}
    \xv_{\mathrm{L}, k+1} = \Amat \xv_{\mathrm{L},k} + \Bmat \uv_{\mathrm{L}, k} + \wv_{\mathrm{L},k},
\end{equation}
while the follower is modeled as 
\begin{equation} \label{x_follower}
    \xv_{\mathrm{F}, k+1} = \Amat \xv_{\mathrm{F},k} + \Bmat \uv_{\mathrm{F}, k} + \wv_{\mathrm{F},k}.
\end{equation}
With $k$ being the discrete time index, $\xv_{\mathrm{L}},~\xv_{\mathrm{F}} \in \mathbb{R}^{n\times 1}$ are the state vectors of the leader and follower, respectively.  $\Amat \in \mathbb{R}^{n \times n}$ and $\Bmat \in \mathbb{R}^{n \times m}$ are constant matrices and $\wv_{\mathrm{L}},~\wv_{\mathrm{F}} \in \mathbb{R}^{n \times 1}$ represent the plant noise, that is Gaussian distributed with zero mean and covariance $\Wmat_{\mathrm{L}}$ and $\Wmat_\mathrm{F}$, respectively. Furthermore, $\uv_{\mathrm{L}},\uv_{\mathrm{F}} \in \mathbb{R}^{m\times 1}$ denote the control inputs for the trajectory following of the leader and for the formation control of the follower, respectively.

Furthermore, let $\hat{\xv}_{\mathrm{L},k}$ be the follower's estimate of the state of the leader. We define 
\begin{subequations}
\begin{align}
\ev_k &= \xv_{\mathrm{F},k} - \Cmat {\xv}_{\mathrm{L},k} - \Bar{\sv}, \label{e}\\
\hat{\ev}_k &= \xv_{\mathrm{F},k} - \Cmat \hat{\xv}_{\mathrm{L},k} - \Bar{\sv} \label{e_hat}
\end{align}
\end{subequations}
as the control deviation and its estimate known to the follower, respectively. $\Cmat$ is a diagonal matrix with elements being either zero or one, depending on which elements of the leader's state the follower is supposed to adapt to. 
The synchronization is accomplished by a proportional, so-called P-controller, such that the control input of the follower is given by 
\begin{equation}
    \uv_{\mathrm{F}, k} = - \Kmat \hat{\ev}_k, \label{u_form}
\end{equation}
where $\Kmat$ is a constant gain matrix. The P-controller is designed based on linear quadratic regulator (LQR) theory (see \cite{anderson2007optimal}). More precisely, we find $\Kmat$ by minimizing the cost function $J = \sum_{k=0}^{\infty} \left( \ev_k^T \Qmat \ev_k + \uv_k^T \Rmat \uv_k\right)$, in which $\Qmat$ and $\Rmat$ denote weighting matrices for the state deviation cost and the control input cost, respectively. The solution of this optimization problem is $\Kmat= (\Bmat^T \Smat \Bmat + \Rmat)^{-1} \Bmat^T \Smat \Amat^T$, where $\Smat$ solves the discrete-time algebraic Riccati equation $\Amat^T \Smat \Amat - \Smat - \Amat^T \Smat \Bmat(\Bmat^T \Smat\Bmat + \Rmat)^{-1}\Bmat^T \Smat \Amat+\Qmat= \Zeromat$. 

From \eqref{e}, we can eventually formulate the control error dynamics, substituting the leader's and follower's state by \eqref{x_leader} and \eqref{x_follower} and using the control law \eqref{u_form}. As a result, we have
\begin{equation} \label{e_dynamics}
\begin{split}
    \ev_{k+1} = ~&\left(\Amat -\Bmat \Kmat\right) \hat{\ev}_k + \Cmat \Amat \left(\hat{\xv}_{\mathrm{L},k} - \xv_{\mathrm{L},k}\right)  + \left(\Amat - \Imat\right) \Bar{\sv}\\ &- \Cmat \Bmat \uv_{\mathrm{L},k} + \wv_{\mathrm{F},k} - \wv_{\mathrm{L},k}.
\end{split}
\end{equation}

\subsection{Communication Model}
The two agents are connected via a wireless MIMO fading channel. Assume that the leader and follower are equipped with $N_\mathrm{L}$ and $N_\mathrm{F}$ antennas, respectively. When an event occurs, the follower sends a request to the leader, which then transmits its current state. The signal received by the follower if an event occurs in the $k$-th time slot is given by 
\begin{equation*}
    \yv_k = \Hmat_k \Fmat_k \xv_{\mathrm{L},k} + \zv_k,
\end{equation*}
where $\Hmat_k \in \mathbb{C}^{N_\mathrm{F} \times N_\mathrm{L}}$ is the channel matrix, $\Fmat_k \in \mathbb{C}^{N_\mathrm{L} \times n}$ is the MIMO precoding matrix and $\zv_k \sim \mathcal{CN}(0,\sigma_z \Imat)$ represents an additive white Gaussian noise (AWGN) vector. Each element of the channel matrix is Gaussian distributed with $\mathcal{CN}(0, 1)$. Moreover, assume that the channel is constant within each time slot. Further assume that the transmission delay is negligible compared to the time constant of the state-space model. 

\subsection{Kalman Filter}
At the follower, a Kalman filter is applied in order to estimate the current state of the leader. Since the measurements (i.e., the signal received from the leader) become available on an event-driven basis, we use the approach from \cite{sinopoli2004kalman}. The state update is calculated based on the system dynamics known at the Kalman filter, while the new observation is taken into account whenever the follower receives an update from the leader. More precisely, the a-priori estimate and the corresponding error covariance are given by 
\begin{align}
    \hat{\xv}_{k|k-1} &= \Amat \hat{\xv}_{k-1|k-1} \\
    \mat{\Sigma}_{k|k-1} &= \mathbb{E}(({\xv}_{k} -\hat{\xv}_{k|k-1}) ({\xv}_{k} -\hat{\xv}_{k|k-1})^H) \nonumber\\
    &=\Amat \mat{\Sigma}_{k-1|k-1} \Amat^H + \Wmat_\mathrm{L} + \Bmat \hat{\Qmat}_\mathrm{u_\mathrm{L}} \Bmat^H,  \label{sigma_apri} 
\end{align}
where $\hat{\Qmat}_\mathrm{u_\mathrm{L}}$ denotes an estimated covariance matrix of the leader's input $\uv_\mathrm{L}$.
The a-posteriori estimate is given by 
\begin{align}
    & &\hat{\xv}_{k|k} &= \hat{\xv}_{k|k-1} + \Gmat_k \left(\yv_k - \Hmat_k \Fmat_K \hat{\xv}_{k|k-1}\right),\\
    & &\mat{\Sigma}_{k|k} &= \left(\Imat - \Gmat_k \Hmat_k \Fmat_k \right) \mat{\Sigma}_{k|k-1}, \label{sigma_apost}
\end{align}
in which
 \begin{equation}
  \Gmat_k = \gamma_{k} \mat{\Sigma}_{k|k-1} \Fmat_k^H \Hmat_k^H \left(\Hmat_k \Fmat_k \mat{\Sigma}_{k|k-1} \Fmat_k^H \Hmat_k^H + \sigma_z^2 \Imat\right)^{-1} \label{kalman_gain}
 \end{equation}
 is the Kalman gain and the binary variable $\gamma_k$ equals one when an event is triggered in time slot $k$ and is equal to zero otherwise. 
 
 \subsection{UAV model}

For performance evaluation of the proposed event-based scheme, we consider a leader-follower formation control of two UAVs. For simplicity, we model the UAV movement in a horizontal plane with coordinates $s_x$ and $s_y$ only, while neglecting the height. The UAVs are represented by a linear state space model according to \eqref{x_leader} and \eqref{x_follower}, which is based on the linearized model from \cite{wang2016dynamics}, \cite{roth2019base}. The state vector is defined as $\xv = \begin{bmatrix}s_x &\dot{s_x} & \vartheta &\dot{\vartheta} & s_y & \dot{s_y} & \phi & \dot{\phi} \end{bmatrix}$, where $\vartheta$ and $\phi$ are the roll and pitch Euler angles, respectively. The control input signal is specified as $\uv = \begin{bmatrix} \theta_1 (n_1^2 - n_2^2) & \theta_2 (n_3^2-n_4^2) \end{bmatrix}^T$, where $\theta_1$ and $\theta_2$ are device-specific parameters and $n_1,\dots,n_4$ represent the rotor speeds. The $\Amat$ and $\Bmat$ matrices are given by
\begin{equation*}
\begin{split}
    \Amat &= e^{\mat{\Tilde{A}} T_s}, \quad \Bmat = \int_0^{T_s} e^{\mat{\Tilde{A}} \alpha} d\alpha \mat{\Tilde{B}},\\
    \mat{\Tilde{A}} &= \begin{bmatrix} \Amat_1 & \Zeromat_{4\times4}\\ \Zeromat_{4\times 4} & \Amat_1 \end{bmatrix}, \quad \Amat_1 = \begin{bmatrix} 0 & 1 & 0 & 0\\ 0 & 0 & g & 0 \\ 0 & 0 & 0 & 1 \\ 0 & 0 & 0 & 0 \end{bmatrix}\\
    \mat{\Tilde{B}} &= \begin{bmatrix} \Bmat_1 & \Zeromat_{4\times1}\\ \Zeromat_{4\times1} & \Bmat_1\end{bmatrix}, \quad \Bmat_1 = \begin{bmatrix} 0 & 0 & 0 & 1\end{bmatrix}^T,
\end{split}
\end{equation*}
where $g$ represents the gravitational acceleration. 

In our simulation, the leader is following a straight line in $s_x$-direction, accelerating from time step $k=0$ up to a constant speed and is then steered in the direction of the follower in order to get around an obstacle on the path (see Fig. \ref{fig:traj}). The control goal is for the follower to keep a constant relative position to the leader while its only knowledge of the leaders position come from the intermittent wireless transmissions and the Kalman filter estimate. 
Furthermore, we assume that prior to the start of the simulation, both agents agree on a common direction of destination and an expected average speed and synchronize the coordinate system such that the $s_x$-axis points in the main direction of flight. This allows the leader to transmit the deviation from the expected $s_x$-coordinate instead of the actual coordinate, which later justifies the assumption that $\| \xv \|$ is bounded.

%% file: content/optimization.tex
We define a quadratic Lyapunov function candidate as $L(\ev_k) = \ev_k^H \Pmat \ev_k$, where $\Pmat$ is a positive semidefinite matrix. The Lyapunov drift is given by $\Delta L_k = L(\ev_{k+1}) - L(\ev_k)$.
For determining the MIMO precoding matrix $\Fmat$, we aim at minimizing the Lyapunov drift \cite{cai2016mimo} subject to a transmission power constraint.
Let $\Tilde{\Amat}=\Amat-\Bmat \Kmat$. Using \eqref{e_dynamics}, the expected value of the Lyapunov drift is given as
\begin{equation}\label{drift}
\begin{split}
    \mathbb{E}\left(\Delta L\right) = &~\hat{\ev}_k^H \Tilde{\Amat}^H \Pmat \Tilde{\Amat} \hat{\ev}_k + \hat{\ev}_k^H \Tilde{\Amat}^H \left(\Pmat+\Pmat^H\right) \left(\Amat - \Imat\right) \Bar{\sv}\\
    & + \Bar{\sv}^H \left(\Amat - \Imat\right)^H \Pmat \left(\Amat - \Imat\right) \Bar{\sv} + \tr \left(\Pmat\left(\Wmat_\mathrm{F} + \Wmat_\mathrm{L}\right)\right.\\ 
    &+\left.\Bmat^H \Cmat^H \Pmat \Cmat \Bmat \Qmat_\mathrm{u_L} + \Amat^H \Cmat^H \Pmat \Cmat \Amat \mat{\Sigma}_{k|k}\right) \\ 
    &- \hat{\ev}_k^H \Pmat \hat{\ev}_k - \tr\left(\Cmat^H \Pmat \Cmat \mat{\Sigma}_{k|k}\right).
    \end{split}
    \raisetag{1\normalbaselineskip}
\end{equation}
Note that in order to guarantee a stable system with the control deviation converging to zero, the Lyapunov drift is required to always be negative. Hence, finding a MIMO precoder that minimizes the expected Lyapunov drift is desirable to achieve satisfactory control performance. Thus, we formulate the following optimization problem
\begin{equation}\label{opt1}
\begin{aligned}
    & \underset{\Fmat_k}{\text{min}}
& &  \mathbb{E}\left(\Delta L_k\right)\\
& \text{s.t.}
& & \tr \left(\Fmat_k^H \Fmat_k\right) \leq \frac{P_\mathrm{max}}{q},\\
\end{aligned}
\end{equation}
where $P_\mathrm{max}$ is the maximum transmit power and $q$ is defined such that $\Vert \xv_{\mathrm{L},k} \Vert ^2 \leq q ,~  \forall k$. 
Considering \eqref{drift}, it is worth noting that only $\mat{\Sigma}_{k|k}$ depends on $\Fmat_k$. Hence, the objective function can be reduced to $\tr((\Amat^H \Cmat^H \Pmat \Cmat \Amat - \Cmat^H\Pmat \Cmat)\mat{\Sigma}_{k|k})$.
Using \eqref{sigma_apost} and \eqref{kalman_gain} and then applying the Woodbury Matrix Identity, the a-posteriori error covariance can be written as
\begin{equation} \label{sigma_kk}
    \begin{split}
        \mat{\Sigma}_{k|k}
        &= \left(\Imat + \frac{1}{\sigma_z^2} \mat{\Sigma}_{k|k-1} \Fmat_k^H \Hmat_k^H \Hmat_k \Fmat_k\right)^{-1} \mat{\Sigma}_{k|k-1} \\
        &= \left(\mat{\Sigma}_{k|k-1}^{-1} + \frac{1}{\sigma_z^2} \Fmat_k^H \Hmat_k^H \Hmat_k \Fmat_k\right)^{-1}.
    \end{split}
\end{equation}
Similar to \cite{cai2016mimo}, we aim at obtaining a closed-form solution.
To this end, we relax the problem utilizing \cite[Theorem~1]{fang1994inequalities} to upper bound the objective function as follows
\begin{equation*}
    \tr \left((\Amat^H \Cmat^H \Pmat \Cmat \Amat - \Cmat^H \Pmat \Cmat)\mat{\Sigma}_{k|k}\right) \leq \lambda_\mathrm{max}(\Mmat) \tr\left( \mat{\Sigma}_{k|k}\right),
\end{equation*}
with $\Mmat = \frac{1}{2}(\Amat^H \Cmat^H (\Pmat+\Pmat^H)\Cmat \Amat - \Cmat^H(\Pmat+\Pmat^H)\Cmat)$ and $\lambda_\mathrm{max}(\Mmat)$ being the maximum eigenvalue of $\Mmat$. Note that inserting \eqref{sigma_kk} is helpful for the derivation of a closed-form solution as we avoid having a matrix product within the trace operator.
Hence, instead of solving \eqref{opt1} we consider the following optimization problem
\begin{equation}\label{opt2}
\begin{aligned}
    & \underset{\Fmat_k}{\text{min}}
& &  \lambda_\mathrm{max}(\Mmat) \tr\left(\left(\mat{\Sigma}_{k|k-1}^{-1} + \frac{1}{\sigma_z^2} \Fmat_k^H \Hmat_k^H \Hmat_k \Fmat_k\right)^{-1}\right)\\
& \text{s.t.}
& & \tr \left(\Fmat_k^H \Fmat_k\right) \leq \frac{P_\mathrm{max}}{q}.\\
\end{aligned}
\end{equation}
Problem \eqref{opt2} is not convex in $\Fmat_k$, but as its structure is similar to the problem studied in \cite{cai2016mimo}, we make use of the technique proposed there in order to show that it is equivalent to a convex problem. First, we define $\Hmat_k = \Umat_k \mat{\Pi}_k \Vmat_k^H$ as the singular value decomposition (SVD) of the channel matrix and $\mat{\Sigma}_{k|k-1} = \Smat_k \mat{\Lambda}_k \Smat_k^T$ as the eigenvalue decomposition of the a-priori estimation error covariance matrix. Assume that the diagonal elements in both $\mat{\Pi}_k$ and $\mat{\Lambda}_k$ are sorted in descending order. 
We further introduce the substitution $\Xmat_k = \mat{\Pi}_k \Vmat_k^H \Fmat_k \Smat_k$. Then, the optimization problem can be rewritten as
\begin{equation}\label{opt3}
\begin{aligned}
  & \underset{\Xmat_k}{\text{min}}
  & & \lambda_\mathrm{max}(\Mmat) \tr \left( \left(\mat{\Lambda}_k^{-1} + \frac{1}{\sigma_z^2} \Xmat_k^H \Xmat_k\right)^{-1} \right)\\
  & \text{s.t.}
& & \tr \left(\mat{\Pi}_k^{-2} \Xmat_k \Xmat_k^H \right) \leq \frac{P_\mathrm{max}}{q}.
\end{aligned}
\end{equation}
Introducing $\mu_k$ as the Lagrange multiplier for the power constraint, the KKT-conditions are formulated as
\begin{align*}
    -\frac{\lambda_\mathrm{max} (\Mmat) }{\sigma_z^2} \Xmat_k \left( \mat{\Lambda}_k^{-1} + \frac{1}{\sigma_z^2} \Xmat_k^H \Xmat_k \right)^{-2} + \mu_k \mat{\Pi}_k^{-2} \Xmat_k = \Zeromat\\
    \mu_k \tr \left(\mat{\Pi}_k^{-2} \Xmat_k \Xmat_k^H \right) = 0\\
    \tr \left(\mat{\Pi}_k^{-2} \Xmat_k \Xmat_k^H \right) - \frac{P_\mathrm{max}}{q} \leq 0\\
    \mu_k \geq 0.
\end{align*}
Note that these are necessary, but not sufficient conditions since \eqref{opt3} is still non-convex. The first equation can be rewritten as 
\begin{equation*}
    \Xmat_k \left( \mat{\Lambda}_k^{-1} + \frac{1}{\sigma_z^2} \Xmat_k^H \Xmat_k \right)^{-2} = \frac{\mu_k \sigma_z^2}{\lambda_\mathrm{max}(\Mmat)} \mat{\Pi}_k^{-2} \Xmat_k.
\end{equation*}
Note, that this equation implies that each row of $\Xmat_k$ corresponding to a non-zero element of $\mat{\Pi}_k$ is a left eigenvector of $\left( \mat{\Lambda}_k^{-1} + \frac{1}{\sigma_z^2} \Xmat_k^H \Xmat_k \right)^{-2}$. From the eigenvalue definition $ \vv \Amat = \lambda \vv$ it follows that $ \vv \Amat^n = \lambda^n \vv$. Hence, we can write
\begin{equation*}
    \Xmat_k \left( \mat{\Lambda}_k^{-1} + \frac{1}{\sigma_z^2} \Xmat_k^H \Xmat_k \right) = \sqrt{\frac{\lambda_\mathrm{max}(\Mmat)}{\mu_k \sigma_z^2}} \mat{\Pi}_k \Xmat_k.
\end{equation*}
When multiplying by $\Xmat_k^H$ from the left, we get
\begin{equation*}
    \underbrace{\Xmat_k^H \Xmat_k }_{\text{sym.}}  \underbrace{ \left( \mat{\Lambda}_k^{-1} + \frac{1}{\sigma_z^2} \Xmat_k^H \Xmat_k \right)}_{\text{sym.}} = \underbrace{\sqrt{\frac{\lambda_\mathrm{max}(\Mmat)}{\mu_k \sigma_z^2}} \Xmat_k^H \mat{\Pi}_k \Xmat_k}_{\text{sym.}}.
\end{equation*}
From the fact that two symmetric matrices commute if their product is also symmetric, we conclude that $\Xmat_k^H \Xmat_k$ and $\mat{\Lambda}_k^{-1}$ commute and are thus simultaneously diagonalizable \cite{horn2012matrix}, meaning that there exists a matrix $\Tmat$ such that $\Xmat_k^H \Xmat_k = \Tmat \diag (\xi_1,\dots,\xi_n) \Tmat^T$ and $\mat{\Lambda}_k^{-1} = \Tmat \diag(\lambda_1^{-1},\dots,\lambda_n^{-1}) \Tmat^T$. Now we can see that for all $\Xmat_k$ that satisfy the KKT-conditions the value of the objective function in \eqref{opt3} only depends on the eigenvalues of $\Xmat_k^H \Xmat_k$, but not on the structure of $\Xmat_k$. Hence, without loss of generality we can assume $\Xmat_k$ to be diagonal. By defining $\Ymat_k = \Xmat_k^H \Xmat_k$, problem \eqref{opt3} is equivalent to 
\begin{equation}\label{opt4}
\begin{aligned}
  & \underset{\Ymat_k}{\text{min}}
  & & \lambda_\mathrm{max}(\Mmat) \tr \left( \left(\mat{\Lambda}_k^{-1} + \frac{1}{\sigma_z^2} \Ymat_k \right)^{-1} \right)\\
  & \text{s.t.}
& & \tr \left(\mat{\Pi}_k^{-2} \Ymat_k \right) \leq \frac{P_\mathrm{max}}{q}.
\end{aligned}
\end{equation}
Note that this problem is convex in $\Ymat_k$. By solving \eqref{opt4} and applying $\Fmat_k = \Vmat_k \mat{\Pi}_k^{-1} \Ymat_k^{1/2} \Smat_k^T$, the solution for the precoding matrix is given by 
\begin{equation}\label{F_opt}
    \Fmat_k^* = \Vmat_k \mat{\Pi}_k^{-1} \left(\left[\sqrt{\frac{\lambda_\mathrm{max}(\Mmat)\sigma_z^2}{\mu_k}} \mat{\Pi}_k - \sigma_z^2 \mat{\Lambda}_k^{-1}\right]^+\right)^{1/2} \Smat_k^T.
\end{equation}
We obtain the Lagrange multiplier $\mu_k$ from water-filling algorithm, such that $\tr\left({\Fmat_k^*}^H \Fmat_k^*\right) = \frac{P_\mathrm{max}}{q}$. Note that \eqref{F_opt} can be interpreted as a water-filling solution as further described in \cite{cai2016mimo} with a dynamic ``water level'' depending on the channel strength and a variable ``seabed level'' that decreases when the estimation error covariance (i.e., transmission urgency) increases.

\subsection{Event-based Communication Algorithm}
The event-triggering policy is inspired by \cite{santos2019self} and consists of two basic principles.
\begin{enumerate}
    \item As long as the expected value of the Lyapunov function at time step $k$ is below a predefined threshold $L_\mathrm{max}$, there is no communication between leader and follower. 
    \item If $\mathbb{E}\left(\ev_k^H \Pmat \ev_k\right)$ exceeds $L_\mathrm{max}$, a state update will be requested by the follower only if the expected Lyapunov drift is not negative.
\end{enumerate}
Mathematically speaking, the precoding matrix is determined as
\begin{equation*}
    \Fmat_k = \begin{cases}
    \Fmat_k^*, &\text{if $\mathbb{E}\left(\ev_k^H \Pmat \ev_k\right) > L_\mathrm{max}$ and $\mathbb{E}\left( \Delta L_k\right) \leq 0$}\\
    \Zeromat, &\text{otherwise,}\\
\end{cases}
\end{equation*}
in which $L_\mathrm{max}$ is a real positive number.

%% file: content/simulation.tex
\begin{figure*}[htb]
    \centering
    \input{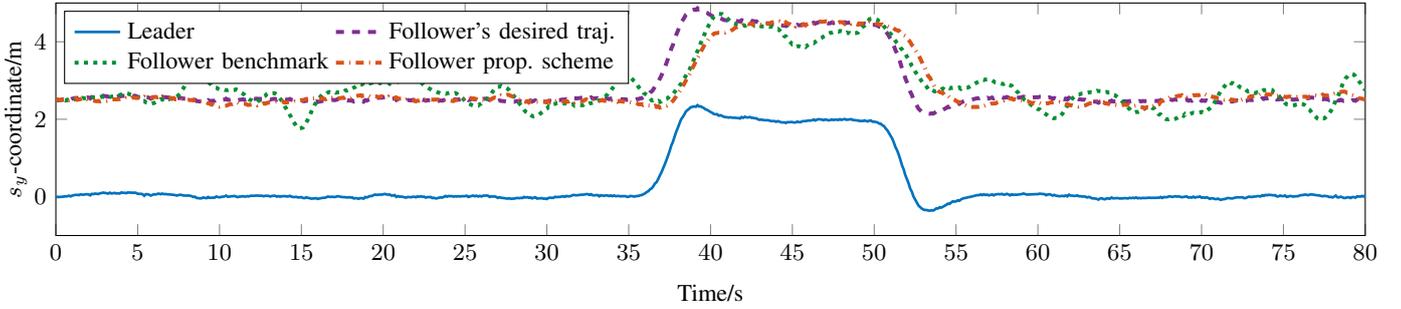}
    \caption{Trajectory of the leader and follower with the proposed communication scheme versus the benchmark scheme. This figure shows the $s_y$-coordinate of the trajectory as a function of time.}
    \label{fig:traj}
\end{figure*}

We compare our proposed event-based scheme with a baseline scheme with periodic transmissions and a general capacity-maximizing water-filling power allocation scheme \cite{cover2012elements}. For a fair comparison, the fixed transmission interval $\Delta T$ of the baseline scheme is chosen to be equal to the average transmission interval of the event-based scheme. The precoding matrix is given by $\Fmat_k = \Vmat_k \mat{\Gamma}_k$, where $\mat{\Gamma}_k = \diag (\gamma_{k,1},\dots,\gamma_{k,n})$ and $\gamma_{k,i} = \left[w-\frac{1}{{\pi}_{k,i}} \right]^+$, such that $\tr(\mat{\Gamma}_k) = \frac{P_\mathrm{max}}{q}$.

The parameters used for the simulation are summarized in Table \ref{tab:param}. Figure \ref{fig:traj} shows the $s_y$-coordinate of the leader's and follower's trajectory with both communication schemes. We observe that the follower's path is more smooth when using our proposed scheme.

\begin{table}[htb]
    \centering
    \begin{tabular}{|c|c||c|c|}
    \hline
        $N_\mathrm{L} = N_\mathrm{F}$ & 8 &
       $T_\mathrm{s}$ & 0.1s \\\hline
       $\Wmat_\mathrm{L} = \Wmat_\mathrm{F}$ & $10^{-5} \cdot \Imat_n$ &
        $\Qmat$ & 10$\cdot\Imat_n$\\ \hline
        $ \Cmat$ & $\diag (1,~1,~ 0,~ 0,~ 1,~ 0,~ 0,~ 0)$ &
       $\Rmat$ & $\Imat_n$\\ \hline
        $\Bar{\sv}$ &  $[0 ~0~ 0~ 0~ 2.5~ 0~ 0~ 0]^T$ &
         $q$ & 3\\ \hline
         $\Pmat$ & $\diag(1,~0,~ 0,~ 0,~ 1,~ 0,~ 0,~ 0)$ & $\Qmat_\mathrm{u}$& 0.3 $\Imat_m$  \\ \hline
    \end{tabular}
    \vspace{2mm}
    \caption{Simulation parameters\\~\\}
    \label{tab:param}
\end{table}

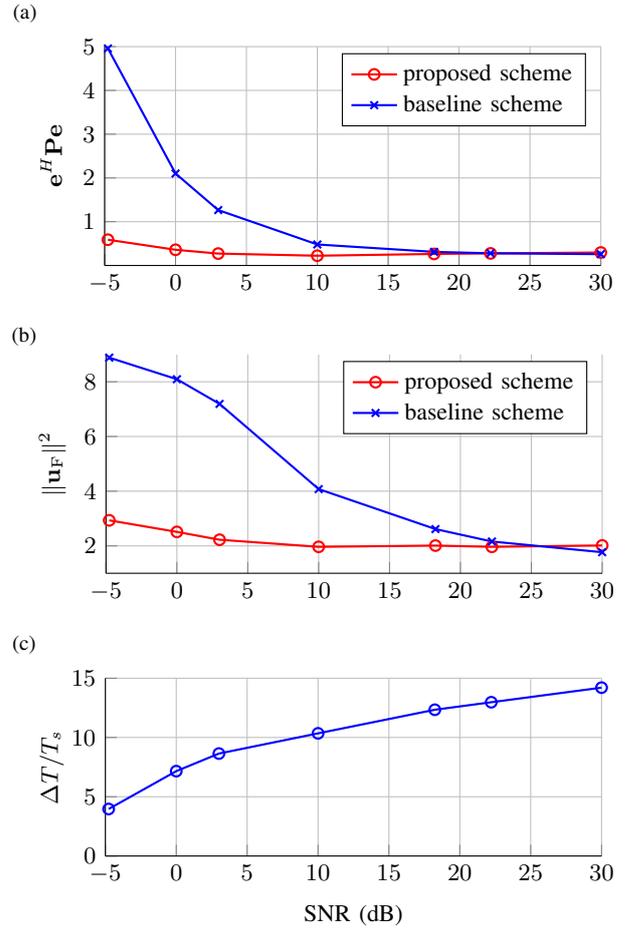
\begin{figure}[htb]
    \centering
     \subfigure[]{\label{fig:SNR_error}
    \input{figures/SNR_error.tex}}
      \subfigure[]{\label{fig:SNR_input}
    \input{figures/SNR_input.tex}}
    \subfigure[]{\label{fig:SNR_vs_T}
    \input{figures/delta_T.tex}}
    \caption{Average control deviation (a), control input power (b) and transmission interval (c) as a function of the SNR with $L_\mathrm{max} = 1$\\~}
\end{figure}

\begin{figure}[htb]
    \centering
    \subfigure[]{\label{fig:e_max_error}
    \input{figures/e_max_error.tex}}
    \subfigure[]{ \label{fig:e_max_input}
    \input{figures/e_max_input.tex}}
    \caption{Average control deviation (a) and control input power (b) as a function of the average transmission time interval with SNR = 15dB.}
\end{figure}
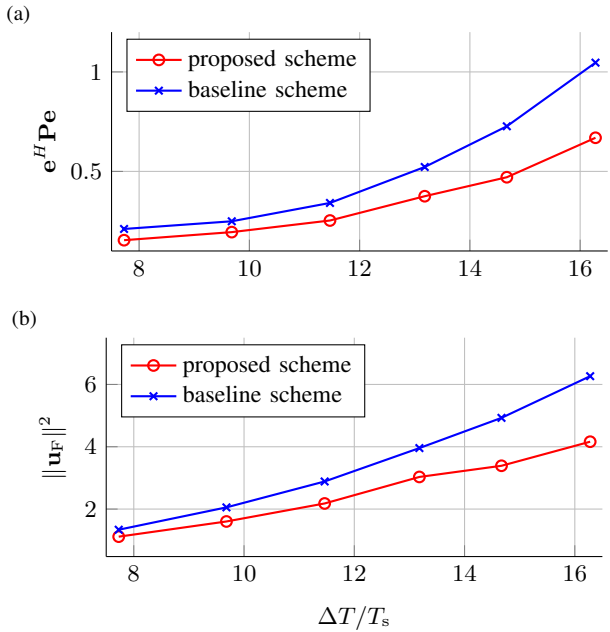

We compare the squared norm of the control deviation and the control input power of the proposed event-based scheme and the periodic transmission baseline scheme averaging over 50 channel and noise realizations. We run the simulations first for different SNR while keeping the other parameters fixed (Fig. \ref{fig:SNR_error}, \ref{fig:SNR_input}). In a second step, we keep a constant SNR changing only the threshold $L_\mathrm{max}$. Note that decreasing $L_\mathrm{max}$ leads to more frequent transmissions and therefore, in Fig. \ref{fig:e_max_error} and \ref{fig:e_max_input}, the $s_x$-axis represents the average transmission interval normalized by the time constant of the state space model.

As shown in Figure \ref{fig:SNR_error}, the proposed scheme clearly outperforms the benchmark scheme especially in the low-SNR regime. At high SNR (> 20dB) both schemes perform equally well. Hence, the proposed scheme allows us to reduce the transmit power without losing performance.
Apart from that, we observe in Figure \ref{fig:SNR_input} that the control input power is significantly lower when using the event-based algorithm. This is due to a more accurate estimate of the Kalman filter, which reduces oscillations of the controller. 
Moreover, Figure \ref{fig:SNR_vs_T} shows that the proposed scheme requires less frequent transmissions on average as the SNR increases. Hence, in the high SNR regime, where the average control deviation converges to a lower bound, the number of transmissions can still be reduced. 

From Figure \ref{fig:e_max_error} it becomes clear that even for constant SNR the average number of transmissions can be reduced in our proposed scheme in order to achieve the same performance as the periodic transmission scheme. The average power of the control input is again lower than in the benchmark scheme (Fig. \ref{fig:e_max_input}).
Hence, the proposed algorithm is more efficient since either the transmit power or the number of transmissions can be reduced without losing performance, while also consuming less control power.

%% file: figures/SNR_error.tex
%
%
\definecolor{mycolor1}{rgb}{0.00000,0.44700,0.74100}%
\definecolor{mycolor2}{rgb}{0.85000,0.32500,0.09800}%

\pgfplotsset{
    tick label style = {font = {\fontsize{9 pt}{9 pt}\selectfont}},
    legend style = {font = {\fontsize{9 pt}{10 pt}\selectfont}}
  }

\begin{tikzpicture}

\begin{axis}[%
width=0.36\textwidth,
height=0.16\textwidth,
at={(0.758in,0.481in)},
scale only axis,
xmin=-5,
xmax=30,
ymin=0,
ymax=5,
xtick={ -5, 0, 5, 10, 15, 20, 25, 30},
ytick={1, 2, 3, 4, 5},
ylabel={$\ev^H \Pmat \ev$},
ylabel style={font={\fontsize{9 pt}{10 pt}\selectfont}},
y label style={yshift=-1.5em},
axis background/.style={fill=white},
axis x line*=bottom,
axis y line*=left,
xmajorgrids,
ymajorgrids,
legend style={legend pos=north east, legend cell align=left, align=left, draw=black}
]
\addplot [color=red, line width=0.8pt, mark=o, mark options={solid, red}]
  table[row sep=crcr]{%
-4.77121254719663	0.588366962775587\\
0	0.358621764732942\\
3.01029995663981	0.271851934329024\\
10	0.222880441482097\\
18.2390874094432	0.263481208180534\\
22.2184874961636	0.275893390421188\\
30	0.296811196858115\\
};
\addlegendentry{proposed scheme}

\addplot [color=blue, line width=0.8pt, mark=x, mark options={solid, blue}]
  table[row sep=crcr]{%
-4.77121254719663	4.95822110400745\\
0	2.10108932462237\\
3.01029995663981	1.26584097565771\\
10	0.479751562287178\\
18.2390874094432	0.310974291040941\\
22.2184874961636	0.278601290925334\\
30	0.258410932248291\\
};
\addlegendentry{baseline scheme}

\end{axis}
\end{tikzpicture}%

%% file: figures/SNR_input.tex
%
%
\definecolor{mycolor1}{rgb}{0.00000,0.44700,0.74100}%
\definecolor{mycolor2}{rgb}{0.85000,0.32500,0.09800}%

\pgfplotsset{
    tick label style = {font = {\fontsize{9 pt}{9 pt}\selectfont}},
    legend style = {font = {\fontsize{9 pt}{10 pt}\selectfont}}
  }

\begin{tikzpicture}

\begin{axis}[%
width=0.36\textwidth,
height=0.16\textwidth,
at={(0.758in,0.481in)},
scale only axis,
xmin=-5,
xmax=30,
ymin=1,
ymax=9,
xtick={ -5, 0, 5, 10, 15, 20, 25, 30},
ylabel={$\| \uv_\mathrm{F} \|^2$},
ylabel style={font={\fontsize{9 pt}{10 pt}\selectfont}},
y label style={yshift=-1.5em},
axis background/.style={fill=white},
axis x line*=bottom,
axis y line*=left,
xmajorgrids,
ymajorgrids,
legend style={legend pos=north east, legend cell align=left, align=left, draw=black}
]

\addplot [color=red, line width=0.8pt, mark=o, mark options={solid, red}]
  table[row sep=crcr]{%
-4.77121254719663	2.93416077443577\\
0	2.51265553549847\\
3.01029995663981	2.22573855548378\\
10	1.966193602268\\
18.2390874094432	2.01170273915917\\
22.2184874961636	1.96610324500917\\
30	2.01687218614813\\
};
\addlegendentry{proposed scheme}

\addplot [color=blue, line width=0.8pt, mark=x, mark options={solid, blue}]
  table[row sep=crcr]{%
-4.77121254719663	8.88236871517894\\
0	8.09348165778043\\
3.01029995663981	7.19120448611116\\
10	4.07679870065456\\
18.2390874094432	2.61527057596921\\
22.2184874961636	2.16070599661571\\
30	1.76943819087264\\
};
\addlegendentry{baseline scheme}

\end{axis}
\end{tikzpicture}%

%% file: figures/delta_T.tex
%
%
\definecolor{mycolor1}{rgb}{0.00000,0.44700,0.74100}%
\definecolor{mycolor2}{rgb}{0.85000,0.32500,0.09800}%

\pgfplotsset{
    tick label style = {font = {\fontsize{9 pt}{9 pt}\selectfont}},
    legend style = {font = {\fontsize{9 pt}{10 pt}\selectfont}}
  }

\begin{tikzpicture}

\begin{axis}[%
width=0.36\textwidth,
height=0.13\textwidth,
at={(0.758in,0.481in)},
scale only axis,
xmin=-5,
xmax=30,
xlabel={SNR (dB)},
xlabel style={font={\fontsize{9 pt}{10 pt}\selectfont}},
ymin=0,
ymax=15,
xtick={ -5, 0, 5, 10, 15, 20, 25, 30},
ylabel={$\Delta T/T_s$},
ylabel style={font={\fontsize{9 pt}{10 pt}\selectfont}},
y label style={yshift=-1.5em},
axis background/.style={fill=white},
axis x line*=bottom,
axis y line*=left,
xmajorgrids,
ymajorgrids,
legend style={legend pos=north east, legend cell align=left, align=left, draw=black}
]
\addplot [color=blue, line width=0.8pt, mark=o, mark options={solid, blue}]
  table[row sep=crcr]{%
-4.77121254719663	3.96532893572487\\
0	7.15912132335614\\
3.01029995663981	8.64224254495559\\
10	10.346127852169\\
18.2390874094432	12.3376050979474\\
22.2184874961636	12.9716227277978\\
30	14.2022413990777\\
};

\end{axis}
\end{tikzpicture}%

%% file: figures/e_max_error.tex
%
%
\definecolor{mycolor1}{rgb}{0.00000,0.44700,0.74100}%
\definecolor{mycolor2}{rgb}{0.85000,0.32500,0.09800}%

\pgfplotsset{
    tick label style = {font = {\fontsize{9 pt}{9 pt}\selectfont}},
    legend style = {font = {\fontsize{9 pt}{10 pt}\selectfont}}
  }

\begin{tikzpicture}

\begin{axis}[%
width=0.36\textwidth,
height=0.16\textwidth,
at={(0.758in,0.481in)},
scale only axis,
xmin=7.5,
xmax=16.5,
xlabel style={font={\fontsize{9 pt}{10 pt}\selectfont}},
ymin=0.1,
ymax=1.2,
ylabel={$ \ev^H \Pmat \ev$},
ylabel style={font={\fontsize{9 pt}{10 pt}\selectfont}},
y label style={yshift=-1em},
axis background/.style={fill=white},
axis x line*=bottom,
axis y line*=left,
xmajorgrids,
ymajorgrids,
legend style={legend pos=north west, legend cell align=left, align=left, draw=black}
]
\addplot [color=red, line width=0.8pt, mark=o, mark options={solid, red}]
  table[row sep=crcr]{%
7.72679202380962	0.154170626993493\\
9.6807368258456	0.194761118844565\\
11.4620683709395	0.253142127960586\\
13.1805883243837	0.375198719106165\\
14.6698822681958	0.470623405906559\\
16.2783497815698	0.668985951738705\\
};
\addlegendentry{proposed scheme}

\addplot [color=blue, line width=0.8pt, mark=x, mark options={solid, blue}]
  table[row sep=crcr]{%
7.72679202380962	0.210594049434994\\
9.6807368258456	0.249248209621762\\
11.4620683709395	0.341394861540213\\
13.1805883243837	0.522095188778468\\
14.6698822681958	0.726720587081192\\
16.2783497815698	1.04658758243309\\
};
\addlegendentry{baseline scheme}

\end{axis}
\end{tikzpicture}%

%% file: figures/e_max_input.tex
%
%
\definecolor{mycolor1}{rgb}{0.00000,0.44700,0.74100}%
\definecolor{mycolor2}{rgb}{0.85000,0.32500,0.09800}%

\pgfplotsset{
    tick label style = {font = {\fontsize{9 pt}{9 pt}\selectfont}},
    legend style = {font = {\fontsize{9 pt}{10 pt}\selectfont}}
  }

\begin{tikzpicture}

\begin{axis}[%
width=0.36\textwidth,
height=0.16\textwidth,
at={(0.758in,0.481in)},
scale only axis,
xmin=7.5,
xmax=16.5,
xlabel={$\Delta T/T_\mathrm{s}$},
xlabel style={font={\fontsize{9 pt}{10 pt}\selectfont}},
ymax=7.5,
ylabel={$\| \uv_\mathrm{F} \|^2$},
ylabel style={font={\fontsize{9 pt}{10 pt}\selectfont}},
y label style={yshift=-1.5em},
axis background/.style={fill=white},
axis x line*=bottom,
axis y line*=left,
xmajorgrids,
ymajorgrids,
legend style={legend pos=north west, legend cell align=left, align=left, draw=black}
]

\addplot [color=red, line width=0.8pt, mark=o, mark options={solid, red}]
  table[row sep=crcr]{%
7.72679202380962	1.11436688057149\\
9.6807368258456	1.60206263821143\\
11.4620683709395	2.18012085196113\\
13.1805883243837	3.03092190327899\\
14.6698822681958	3.38959311305672\\
16.2783497815698	4.16214472456361\\
};
\addlegendentry{proposed scheme}

\addplot [color=blue, line width=0.8pt, mark=x, mark options={solid, blue}]
  table[row sep=crcr]{%
7.72679202380962	1.33722756464484\\
9.6807368258456	2.05443527707731\\
11.4620683709395	2.8857280283778\\
13.1805883243837	3.95783653677944\\
14.6698822681958	4.92737181023905\\
16.2783497815698	6.26520390786161\\
};
\addlegendentry{baseline scheme}

\end{axis}
\end{tikzpicture}%

%% file: content/conlusion.tex
In this paper, we studied a leader-follower formation control problem with one leading and one following agent, each equipped with multiple antennas. While the leader follows a given path, which is unknown to the follower, the follower is supposed to move along with a constant relative position to the leader. An update on the current leader's position may be requested by the following agent via a MIMO point-to-point channel. An event-based MIMO communication scheme was derived based on Lyapunov drift theory. For a performance analysis, we considered a UAV formation control setting. Compared to a benchmark scheme with periodic transmissions and conventional capacity-maximizing water-filling, the proposed scheme performs better with equal transmit power while consuming less energy on the control signal. Furthermore, when using our proposed scheme, the average number of transmissions can be reduced without a performance loss.